\documentclass[aps,pra,twocolumn,superscriptaddress,longbibliography,reprint,floatfix]{revtex4-2}

\usepackage{graphicx}
\usepackage{comment}
\DeclareUnicodeCharacter{2212}{\uniminus}
\usepackage{float}
\usepackage{subfiles}
\usepackage{mathtools}  
\usepackage{amssymb}    
\usepackage{dirtytalk}
\usepackage{braket}
\usepackage{siunitx}
\usepackage{tikz}

\usepackage{stackengine}
\newcommand{\op}[1]{\mathop{}\!\ensurestackMath{\stackon[-.95ex]{%
  {#1}}{\smash{{\hat{}}}}}}
\newcommand{\opdag}[1]{\mathop{}\!\op{#1}^{\dag}}

\setcitestyle{super}

\begin{document}


\title{Entanglement Witnesses of Condensation for Enhanced Quantum Sensing}

\author{Lillian I. Payne Torres}

\author{Irma Avdic}
\author{Anna O. Schouten}
\author{Olivia C. Wedig}
\author{Gregory S. Engel}
\email{gsengel@uchicago.edu}

\author{David A. Mazziotti}
\email{damazz@uchicago.edu}

\affiliation{Department of Chemistry and The James Franck Institute, The University of Chicago, Chicago, IL 60637}%

\date{Submitted November 21, 2025}



\begin{abstract}

Quantum phenomena such as entanglement provide powerful resources for enhancing classical sensing. Here, we theoretically show that collective entanglement of spin qubits, arising from a condensation of particle-hole pairs, can strongly amplify transitions between ground and excited spin states, potentially improving signal contrast in optically detected magnetic resonance.  This collective state exhibits an $\mathcal{O}(\sqrt{N})$ enhancement of the transition amplitude with respect to an applied microwave field, where $N$ is the number of entangled spin qubits. We computationally realize this amplification using an ensemble of $N$ triplet spins with magnetic dipole interactions, where the largest transition amplitudes occur at geometries for which the condensation of particle-hole pairs is strongest. This effect, robust to noise, originates from the concentration of entanglement into a single collective mode, reflected in a large eigenvalue of the particle-hole reduced density matrix---an entanglement witness of condensation analogous to off-diagonal long-range order, though realized here in a finite system. These results offer a design principle for quantum sensors that exploit condensation-inspired entanglement to boost sensitivity in spin-based platforms.

\begin{figure}[H]
\centering

\end{figure}
\end{abstract}

\maketitle

\section{Introduction}

Quantum sensors have the potential to lead to unprecedented advances in condensed matter physics and the biological sciences by harnessing the power of quantum mechanics to offer non-invasive, quantitative, nanoscale resolution measurements of physical parameters such as magnetic fields,\cite{balasubramanian_nanoscale_2008,degen_scanning_2008,glenn_high-resolution_2018} electric fields,\cite{dolde_electric-field_2011,bian_nanoscale_2021,neumann_high-precision_2013}
and temperature.\cite{neumann_high-precision_2013,kucsko_nanometre-scale_2013,toyli_fluorescence_2013}
Optically addressable spin qubits are a highly promising modality for quantum sensing, demonstrating high-fidelity initialization, control, and readout of quantum states. In particular, solid-state defect centers such as the nitrogen-vacancy (NV) center in diamond have attracted much interest due to their long coherence times and high-resolution optical readout.\cite{doherty_nitrogen-vacancy_2013,awschalom_quantum_2018,glenn_high-resolution_2018,schirhagl_nitrogen-vacancy_2014,matsuzaki_optically_2016,gruber_scanning_1997} Other promising candidates include molecular systems and metal-organic frameworks, which have the added benefits of controllable spatial resolution and high tuneability,\cite{kultaeva_atomic-scale_2022,yamabayashi_scaling_2018,bayliss_optically_2020} as well as fluorescent proteins, which are genetically expressible and easily integrated into biological systems.\cite{feder_fluorescent-protein_2024} However, despite recent advancements, the practical realization of precise quantum measurements in noisy biological environments remains an open challenge.

Recent efforts have sought to improve sensitivity in quantum sensing both by increasing the number of sensing qubits and by attempting to extend coherence lifetimes.\cite{rizzato_extending_2023,yamamoto_extending_2013,wood_long_2022,bauch_decoherence_2020,barry_sensitivity_2020} However, the sensitivity achievable with systems of uncorrelated qubits is fundamentally limited by the standard quantum limit, $\mathrm{\delta\theta \geq \frac{1}{\sqrt{N}}}$, which defines the minimum uncertainty possible with increasing number of qubits $N$.\cite{giovannetti_quantum-enhanced_2004,giovannetti_quantum_2006,giovannetti_advances_2011} This limit can be surpassed through the use of entangled qubits, which collectively accumulate information about the physical parameter being sensed and can lead to an up to $\sqrt{N}$-fold improvement in measurement precision, saturating the so-called Heisenberg limit $\mathrm{\delta\theta \geq \frac{1}{N}}$.\cite{giovannetti_quantum-enhanced_2004,giovannetti_quantum_2006,giovannetti_advances_2011} This makes the realization of entangled states of sensing qubits a central goal in the field of quantum sensing.\cite{huang_entanglement-enhanced_2024,zaiser_enhancing_2016} Experimental progress to date has utilized systems such as trapped ions,\cite{roos_complete_1980,monz_14-qubit_2011,strobel_fisher_2014,bohnet_quantum_2016,gilmore_quantum-enhanced_2021} cavity-confined atoms,\cite{leroux_implementation_2010,zhang_detuning-enhanced_2015,li_collective_2022,greve_entanglement-enhanced_2022} and Bose-Einstein condensates;\cite{zou_beating_2018,riedel_atom-chip-based_2010,ockeloen_quantum_2013} however, such systems rely on cryogenic temperatures and are generally unsuitable for biosensing applications. The search is ongoing for a quantum sensing platform capable of harnessing the power of quantum entanglement to achieve high-sensitivity measurements in a practical context.

In this work, we show that an entanglement witness of condensation in finite ensembles of $N$ spin qubits reveals the concentration of particle-hole entanglement into a single collective mode that drives an $\mathcal{O}(\sqrt{N})$ enhancement in spin-transition amplitudes.  The entanglement witness---a large eigenvalue of the particle-hole reduced density matrix---is a form of off-diagonal long-range order (ODLRO), albeit realized here in a finite system.  We illustrate this mechanism in an interacting array of triplet spins with magnetic dipole interactions, where the entanglement witness and the transition amplitudes are jointly maximized at specific geometries, revealing an entanglement-dependent pathway for controlling the response of spin-based quantum sensors.  This increase in spin-transition amplitudes can translate into greater spin-dependent optical contrast in optically detected magnetic resonance.  Similar particle-hole entangled states have demonstrated resilience to environmental noise,\cite{schouten_exciton-condensate-like_2023, schouten_exciton-condensate-like_2025} which is highly advantageous for high-precision quantum sensing, including in biological environments. Together, these results establish a general design principle for improving the sensitivity of spin qubits with potential applications to emerging quantum sensor platforms such as molecular arrays and fluorescent proteins.

\section{Theoretical Background}
\subsection*{Enhanced Collective States}

To improve signal to noise ratio and thus increase sensitivity, one can enhance the physical process responsible for the signal. The quantum sensing capabilities of many spin qubits, such as NV centers, are based around optical manipulation and readout of the qubit spin states. One widely employed set of techniques is optically detected magnetic resonance (ODMR), wherein the population of the qubit spin states is read out from the spin-dependent fluorescence spectrum.\cite{davies_optically-detected_1976,carbonera_optically_2009,abe_tutorial_2018,bayliss_optically_2020,jelezko_read-out_2004,zhang_little_2018} A significant dip in the fluorescence intensity will occur for microwave frequencies resonant with the spin state transitions, allowing readout of any applied field due to the proportionality between applied field strength and the splitting of the spin sub-levels.\cite{yoon_identifying_2025,zhang_little_2018} The signal is thus inherently dependent on the strength of the spin-state transitions. The strength of the transition can be thought of as the transition amplitude between the ground spin state and some higher-energy spin state, $\langle \Psi_\mathrm{e} | \op{T} | \Psi_\mathrm{g} \rangle$, where $\hat{T}$ is the transition operator for the process.

One avenue for amplification of these transitions is through the creation of an \textit{enhanced collective state}, a state for which there exists some transition operator having a large transition amplitude between that state and the ground state:\cite{garrod_particle-hole_1969}
\begin{equation}
\langle \Psi_\mathrm{e} | \op{T} | \Psi_\mathrm{g} \rangle >> 1.
\end{equation}
The origins of such a state can be understood via the particle-hole reduced density matrix, which is defined generally as
\begin{equation}
{}^2G^\mathrm{i,j}_\mathrm{k,l} = \bra{\Psi}   \opdag{a}_\mathrm{i}\hat{a}^{}_\mathrm{j}\opdag{a}_\mathrm{l}\hat{a}^{}_\mathrm{k} \ket{\Psi},
\label{Gmat}
\end{equation}
where $\ket{\Psi}$ is an $N$-particle wavefunction, and $\opdag{a}_\mathrm{i}$ and $\hat{a}^{}_\mathrm{i}$ are the creation and annihilation operators for each spin state.\cite{garrod_particle-hole_1969,safaei_quantum_2018,kohn_two_1970}  Examining the particle-hole RDM for a particular excited spin state $\ket{\Psi_\mathrm{e}}$, we expand the expression for the matrix elements as
\begin{equation}
{}^2G^\mathrm{i,j}_\mathrm{k,l}=\sum_\mathrm{s} \bra{\Psi_\mathrm{e}}   \opdag{a}_\mathrm{i}\hat{a}^{}_\mathrm{j} \ket{\Psi_\mathrm{s}}\bra{\Psi_\mathrm{s}}\opdag{a}_\mathrm{l}\hat{a}^{}_\mathrm{k} \ket{\Psi_\mathrm{e}}
\end{equation}
where $\ket{\Psi_\mathrm{s}}$ are the particle-hole states. In order to retain only on the particle-hole states created by single-spin transitions, we subtract the components corresponding to a state-to-state projection to create the modified particle-hole RDM:
\begin{equation}
{}^2\Tilde{G}^\mathrm{i,j}_\mathrm{k,l}=\sum_\mathrm{s \neq e} \bra{\Psi_\mathrm{e}}   \opdag{a}_\mathrm{i}\hat{a}^{}_\mathrm{j} \ket{\Psi_\mathrm{s}}\bra{\Psi_\mathrm{s}}\opdag{a}_\mathrm{l}\hat{a}^{}_\mathrm{k} \ket{\Psi_\mathrm{e}},
\end{equation}
where the $\ket{\Psi_\mathrm{s}}$ corresponds to single-spin transitions from $\ket{\Psi_\mathrm{e}}$.  The eigenvalues of this matrix correspond to the occupations of the particle-hole states. In the non-interacting limit, the eigenvalues will be either zero or one.  However, in the presence of correlation, larger-than-one eigenvalues can form.  The emergence of a large eigenvalue $\mathcal{O}(\sqrt{N})$ in this matrix corresponds to the condensation of the particle-hole pairs into a single collective, entangled state, known as ODLRO.\cite{safaei_quantum_2018, garrod_particle-hole_1969}


To understand why transitions into such a state are enhanced, we define a one-body transition operator $\op{T}$ from the eigenvector corresponding to the large eigenvalue
\begin{equation}
\op{T}=\sum_\mathrm{ij} T_\mathrm{ij}\opdag{a}_\mathrm{i}\hat{a}^{}_\mathrm{j};
\end{equation}
where $T_\mathrm{ij}$ contains the components of the eigenvector. The large eigenvalue $\lambda$ can thus be defined as
\begin{align}
 \lambda &= \sum_\mathrm{ijkl} T^\mathrm{i}_\mathrm{j} {}^2\Tilde{G}^\mathrm{i,j}_\mathrm{k,l} T^\mathrm{l^*}_k \\
& =\sum_\mathrm{s \neq e}\sum_\mathrm{ijkl} T^\mathrm{i}_j T^\mathrm{k^*}_l \bra{\Psi_\mathrm{e}}   \opdag{a}_i\hat{a}^\mathrm{}_j
\ket{\Psi_\mathrm{s}}\bra{\Psi_\mathrm{s}}\opdag{a}_l\hat{a}^\mathrm{}_k \ket{\Psi_\mathrm{e}} \\
& =\sum_\mathrm{s \neq e} \bra{\Psi_\mathrm{e}}   \op{T} \ket{\Psi_\mathrm{s}}\bra{\Psi_\mathrm{s}}\op{T} \ket{\Psi_\mathrm{e}}
\end{align}
The summation can be truncated, assuming that this transition overlaps strongly with the ground state,
\begin{align}
     \lambda &\approx \bra{\Psi_\mathrm{e}}   \op{T} \ket{\Psi_\mathrm{g}}\bra{\Psi_\mathrm{g}}\op{T} \ket{\Psi_\mathrm{e}} \\
     \lambda^\mathrm{\frac{1}{2}} &\approx \bra{\Psi_\mathrm{e}}   \op{T} \ket{\Psi_\mathrm{g}} .
     \label{eq:lambdarad}
\end{align}
We can thus see that the transition amplitude, which is intrinsically linked to the eigenvalue of the modified particle-hole RDM, will be large for a state with strong correlations between different spin transitions, reflected in a large $\lambda$ value. As correlation increases, the transition amplitude will increase concomitantly with $\lambda$ until $\lambda$ saturates the maximal value of $N/2$,\cite{garrod_particle-hole_1969}, corresponding to the maximal occupation of the collective state and thus the maximal entanglement limit. For this amplitude enhancement to be practically realized, two conditions must be met: the transition operator must (1) overlap strongly with the ground state such that a transition between the ground state and the enhanced collective state is possible and (2) correspond to a physically relevant process such as a transition dipole.

The large eigenvalue of the particle-hole RDM has frequently been discussed in the context of exciton condensation, where it serves as a characteristic signature of the presence of an exciton condensate.\cite{payne_torres_molecular_2025, payne_torres_molecular_2024, schouten_potential_2023, sager_beginnings_2022, schouten_exciton_2021, safaei_quantum_2018} In this context the eigenvalues of the particle-hole RDM are the populations of the excitonic states, and an eigenvalue larger than one indicates more than one exciton occupying the same quantum state, or the beginnings of an exciton condensate. Similarly, we can view a large eigenvalue in the particle-hole RDM for our spin qubit system as corresponding to the occupation of a strongly correlated collective spin state and the beginnings of a kind of quantum condensate state of spin qubits.


\subsection*{Model}
\begin{figure}
    \centering
    \includegraphics[width=8cm]{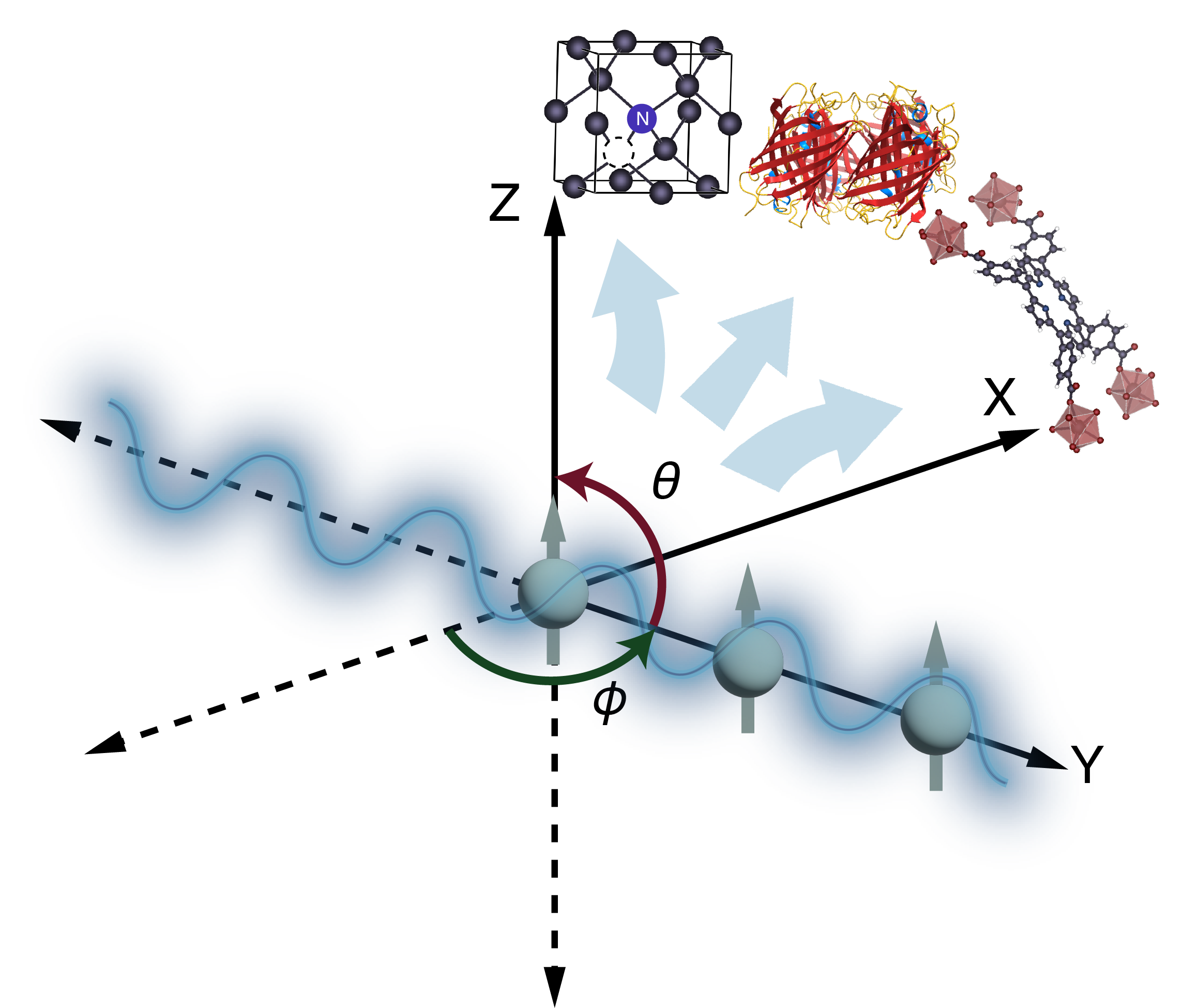}
    \caption{Schematic of model system. Three qubits, shown in gray, are spaced equidistant along the Y axis with spins aligned along the Z axis. The microwave, illustrated in blue, is applied along the Y axis (wavelength not to scale). Inter-qubit spacing can be scaled, and qubits can be rotated about the polar ($\theta$) and azimuthal ($\phi$) axes. While the zero field splitting parameter used in the model Hamiltonian corresponds to the NV center parameter in this work, similar models could easily be constructed to represent fluorescent protein or molecular lattice-based spin qubits.}
    \label{fig:model}
\end{figure}

Our model system consists of an ensemble of \textit{N} triplet spins aligned along the Z axis in a \textit{d}-dimensional array with magnetic dipole interactions. A basic schematic of this system is given in Fig.~\ref{fig:model}. An individual isolated spin can be described by the following spin Hamiltonian:
\begin{equation}
\op{H}^\mathrm{\mathrm{(i)}} = D(\op{S}_\mathrm{Z}^2-\frac{1}{3}\op{S}^2),
\label{eq:Hi}
\end{equation}
where \textit{D} is the axial zero-field splitting parameter and $\op{S}=(\op{S}_\mathrm{X}, \op{S}_\mathrm{Y}, \op{S}_\mathrm{Z} )$ are the triplet spin operators.\cite{doherty_theory_2012} In this work, we use a value of $D =2.87$~GHz, corresponding to the zero-field splitting parameter for a NV center at ambient temperature and pressure.\cite{gruber_scanning_1997} The eigenvectors of this Hamiltonian are the spin sub-levels $\ket{m_\mathrm{s}=0}$ and $\ket{m_\mathrm{s}=\pm1}$, where the latter are degenerate in the absence of a perturbative field.\cite{harrison_optical_2004} In the presence of a perturbative field such as a magnetic or electric field, this degeneracy will be broken and the splitting between the states will be proportional to the strength of the field.\cite{zhang_little_2018}
The total Hamiltonian for the interacting triplet system is:
\begin{equation}
\op{H}^\mathrm{\mathrm{N}}_\mathrm{\mathrm{interacting}} = \sum_\mathrm{i=1}^\mathrm{N}\op{H}^\mathrm{\mathrm{(i)}} + \sum_\mathrm{i<j}^\mathrm{N}\op{H}_\mathrm{\mathrm{dipole}}^\mathrm{\mathrm{(ij)}},
\label{eq:Htot}
\end{equation}
where $\op{H}^\mathrm{\mathrm{(i)}}$ are the Hamiltonians for the individual triplet spins and $\op{H}^\mathrm{(ij)}_\mathrm{dipole}$ describes their mutual magnetic dipole interaction.\cite{van_oort_low-field_1990,hanson_polarization_2006} The Hamiltonian for the magnetic dipole interaction between the triplets is written as:
\begin{equation}
\op{H}^\mathrm{(ij)}_\mathrm{\mathrm{dipole}} =\frac{\hbar\gamma_\mathrm{el}^2}{|R_\mathrm{ij}|^3}\left[\op{S}^\mathrm{i}\cdot\op{S}^\mathrm{j}-3\left(\op{S}^\mathrm{i}\cdot\frac{\op{R}_\mathrm{ij}}{|R_\mathrm{ij}|}\right)\left(\op{S}^\mathrm{j}\cdot\frac{\op{R}_\mathrm{ij}}{|R_\mathrm{ij}|}\right)\right],
\label{eq:Hdip}
\end{equation}
where $\op{S}^\mathrm{i}$ and $\op{S}^\mathrm{j}$ are the spin operators of each triplet, $\gamma_\mathrm{e}$ is the electron gyromagnetic ratio, $\op{R}_\mathrm{ij}$ is the displacement vector between the two triplets, and $|R_\mathrm{ij}|$ is the magnitude of the displacement vector.

As the spin state transitions in ODMR are induced via microwave radiation, the  physically relevant transition operator in this context is the microwave-interaction Hamiltonian:\cite{dreau_avoiding_2011}
\begin{equation}
\op{T}=\op{H}_\mathrm{\mathrm{micro}}= \gamma_\mathrm{e}B_1\cos(\omega t)\left[\op{S}_\mathrm{X}+\op{S}_\mathrm{Y}\right],
\label{eq:Hmicro}
\end{equation}
where $B_1$ is the microwave magnetic field strength and $\omega$ is the microwave frequency. In the rotating frame this is
\begin{equation}
\op{H}_\mathrm{micro}=-\omega\op{S}_\mathrm{Z}+\gamma_\mathrm{e}B_1\left[\cos(\phi)\op{S}_\mathrm{X}+\sin(\phi)\op{S}_\mathrm{Y}\right],
\label{eq:Hmicrorot}
\end{equation}
with the corresponding transition amplitude $A$
\begin{equation}
A = \langle \Psi_\mathrm{e} | \op{H}_\mathrm{\mathrm{micro}} | m_\mathrm{s}=0, m_\mathrm{s}=0, m_\mathrm{s}=0 \rangle.
\label{eq:transamp}
\end{equation}
All transition amplitudes discussed in the following have been normalized such that the amplitude in the non-interacting limit is set to 1~a.u.

\section{Results}
\section*{Sensitivity Enhancement from Spin Qubit Condensation}

\begin{figure}

\includegraphics[width=\linewidth, height=\textwidth, keepaspectratio]{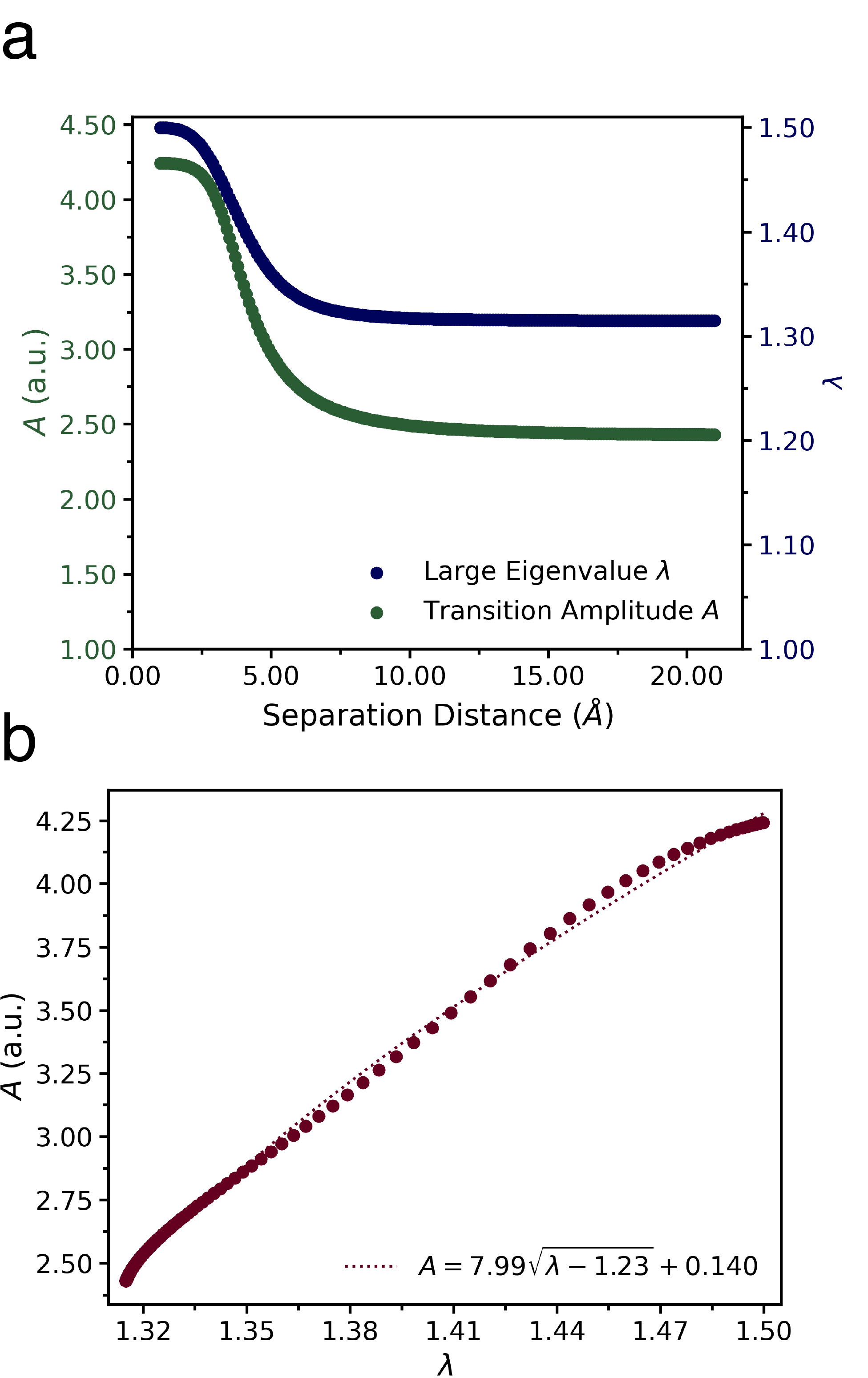}

\caption{(a) Largest transition amplitudes (green) and $\lambda$ value for the corresponding excited spin state (blue) for three centers spaced at increasing distances from each other along the axis of microwave propagation. (b) Transition amplitude plotted against $\lambda$ for three centers with increasing interaction strengths spaced at decreasing distances from each other along the axis of microwave propagation. The best-fit curve is obtained via nonlinear least-squares fit, and shows a square-root dependence of transition amplitude on $\lambda$, as predicated by equation (\ref{eq:lambdarad}) .}
\label{fig:distscale}
\end{figure}

To determine if our interacting spin qubit model can manifest ODLRO in the particle-hole RDM, resembling an enhanced collective state, we first calculate the transition amplitudes for transitions between the lowest energy spin state and each of the higher-energy spin states for a system of three qubits with an inter-qubit separation of 5.125~{\AA}. We find that only the transition from the ground spin state to the first excited spin state has a non-zero amplitude. We then calculate the eigenvalues of the particle-hole RDM for that first excited spin state, finding the value of the large eigenvalue entanglement witness to be 1.36, indicating the presence of spin-state particle-hole entanglement. To investigate the connection between the large eigenvalue and the transition amplitude, and understand if the presence of the entangled collective state is indeed leading to transition enhancement, we calculate the eigenvalues of the particle-hole RDM for the first excited spin state and the transition amplitude for the ground to first excited spin state transition for a range of inter-qubit separations from $2.0$ to $22.0$ {\AA}. This distance range is chosen to be representative of the potential inter-qubit distances in solid-state, molecular, or fluorescent protein-based spin qubit systems. These results are plotted in Fig.~\ref{fig:distscale}a. We find that the large eigenvalue and the transition amplitude decay and then plateau concurrently with increasing distance. The value of $\lambda$ decreases with increasing distance, suggesting that the dipole interaction is responsible for the emergence of ODLRO, and as the dipole interaction strength decreases with increasing distance, the strong correlation dissipates. The fact that $A$ decays concurrently with $\lambda$ supports the connection between the entangled state and the transition amplitude and suggests the presence of an enhanced collective state and corresponding enhanced transition amplitude. Plotting the large eigenvalue against the transition amplitude for each distance results in a square-root relationship between $\lambda$ and $A$, as predicted in our earlier derivation (Fig. \ref{fig:distscale}b).

\begin{figure}
\label{fig:increasequb_amp}
    \includegraphics[width=\linewidth, height=\textwidth, keepaspectratio]{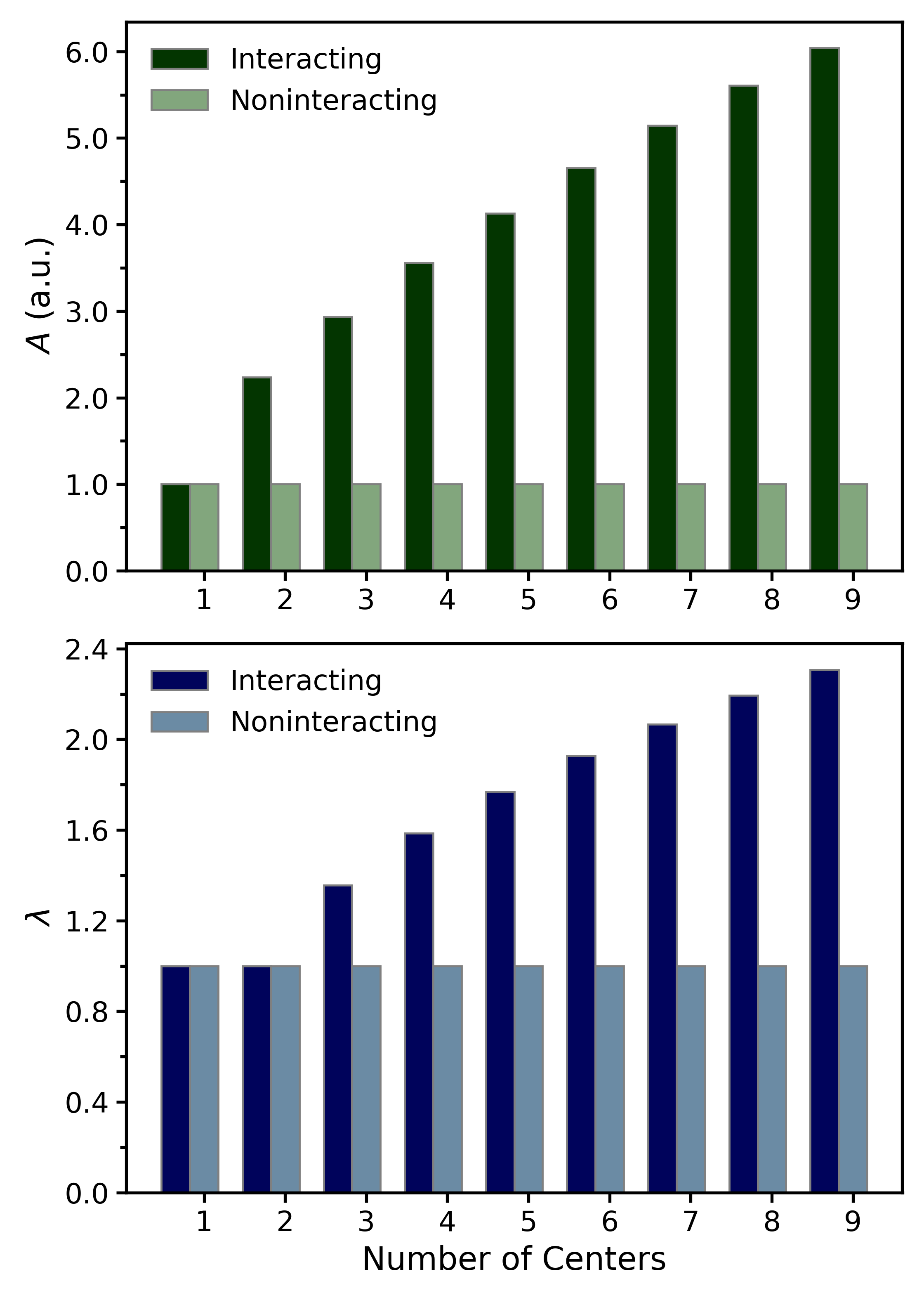}
\caption{The (top) largest transition amplitude and (bottom) $\lambda$ value for the corresponding excited spin state for systems with increasing numbers of spins, with and without dipole interactions. Spins are spaced along the Y axis with 5.125 {\AA} separation. For the non-interacting case, the maximum transition amplitudes and $\lambda$ values are completely unchanging with increasing numbers of qubits. In the presence of dipole interactions, both the transition amplitudes and the values of $\lambda$ display an approximately radical dependence on the number of qubits. }
\label{fig:increasequb}
\end{figure}

\subsection{Increasing System Size}

As mentioned previously, increasing the number of sensing qubits is a fundamental pathway for enhancing signal and thus sensitivity in quantum sensing.\cite{avdic_quantum_2023,zhou_achieving_2018,giovannetti_quantum-enhanced_2004,giovannetti_quantum_2006} It is thus desirable to understand how the signal enhancement from the presence of an enhanced collective state is impacted by increasing qubit number, and how this compares to the effect of increasing qubit number for unentangled sensors. To this end, we calculate $\lambda$ and $A$ for systems with as many as 9 spin qubits, both with dipole interactions (as in the previous calculations) and without such interactions. These results are plotted in Fig.~\ref{fig:increasequb}. We find that when dipole interactions are present, increasing the number of spins causes an increase in both $\lambda$ and the transition amplitude, with an approximately square-root dependence on the number of qubits in each case. However, in the non-interacting case, neither the large eigenvalue nor the transition amplitude increases with the number of qubits. This means that, while there is still the possibility for signal increase with increasing numbers of qubits for non-interacting systems due to the combination of signals from multiple degenerate transitions, there is no fundamental enhancement of the transitions themselves. However, in the case of an interacting system with ODLRO, increasing the number of sensing qubits increases the size of the enhanced collective state, leading to a larger $\lambda$ and thus greater transition amplitude enhancement.

\subsection{2D Systems}

\begin{figure}
\includegraphics[width=\linewidth, keepaspectratio]{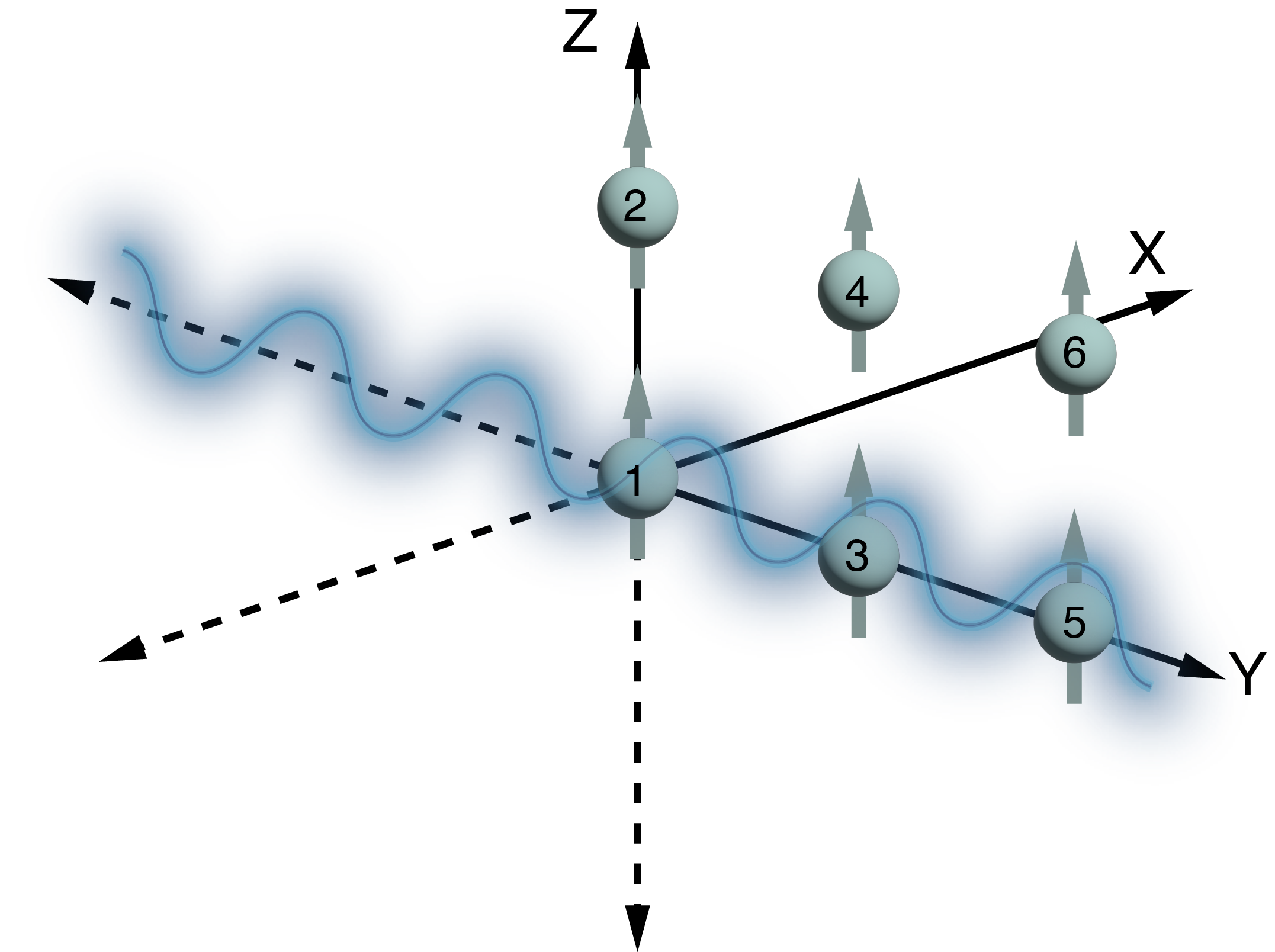}
\caption{Schematic of model system for probing 2D interactions. Spin qubits are spaced equidistant along the Y axis, with additional row of qubits added offset 5.125 {\AA} in Z (or X) direction. The microwave, illustrated in blue, is applied along the Y axis (wavelength not to scale).}
\label{fig:zy_model}
\end{figure}

While in the above example we increased the number of qubits in a 1D chain along the Y axis, precise alignments of spin qubits in a 1D chain would likely be challenging in an experimental context. To understand the effect of additional qubits not aligned in a 1D chain, and thus the addition of interactions in a direction other than that of microwave propagation, we calculate $A$ and $\lambda$ for increasing numbers of qubits in two dimensions. A second layer of qubits is added displaced 5.125 {\AA} along the Z axis or the X axis. These arrangements will be referred to as ZY and XY respectively. A diagram showing the arrangement of qubits and the order in which they are added is provided in Fig.~\ref{fig:zy_model} for the ZY arrangement.

\begin{figure}
\includegraphics[width=\linewidth, height=\textwidth, keepaspectratio]{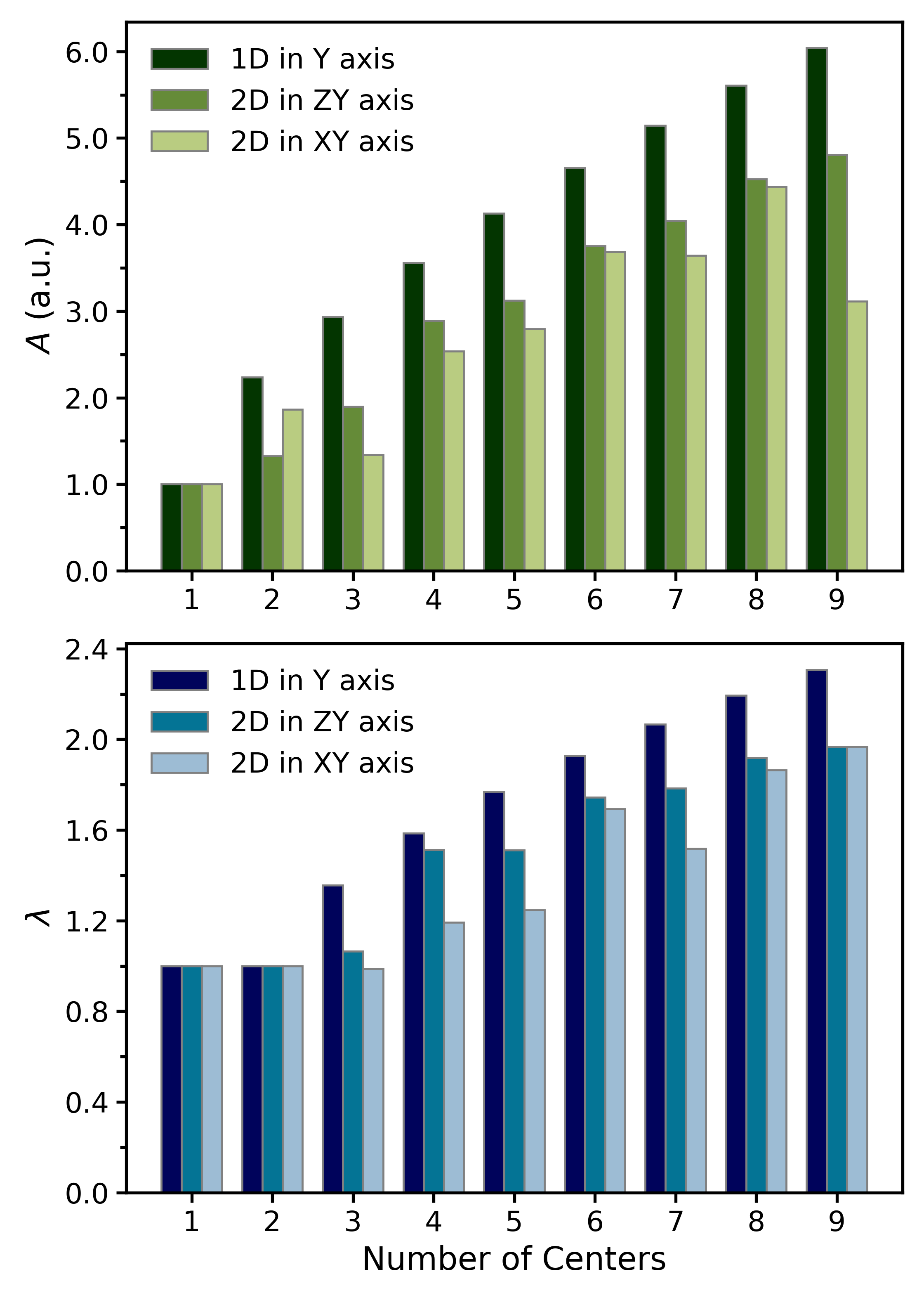}
\caption{(top) The largest transition amplitude for systems with increasing numbers of spins for the ZY and XY arrangements. (bottom) The $\lambda$ value corresponding to the largest transition  $A$ and $\lambda$ values for a 1D chain of spins spaced along the Y axis with 5.125 {\AA} separation are included for comparison.}
\label{fig:center_zy}
\end{figure}

Plots of $A$ and $\lambda$ for increasing numbers of qubits with additional 2D interactions added are shown in Fig. \ref{fig:center_zy} top and bottom for both the ZY and XY arrangements. These results demonstrate that adding additional qubits along a new axis does not increase the size of the enhanced collective state in the same way that adding new qubits in a 1D chain along the axis of microwave propagation does. While the size of the collective state in the ZY case does slowly increase, it only truly increases as new qubits are added in the microwave propagation direction---it takes six qubits to reach approximately the same $A$ and $\lambda$ values exhibited by three qubits in the 1D chain. The size of the collective state in the XY case does generally increase with the number of qubits, but displays fluctuations that are absent in the 1D or 2D ZY cases. The presence of these fluctuations suggests that additional interactions in the X direction cause some amount of disruption to the enhanced collective state, although not enough to fully quench the entangled state. In general, these results imply that it is the interactions along the axis of microwave propagation that are most important for the generation of a robust enhanced collective state, and alignment of qubits along this axis should be maximized. However, it is promising that interactions along other axes do not always appear to quench the enhanced collective state, as a collective state is still present ($\lambda$ and $A  > 1$) in the presence of off-axis interactions, and still grows with increasing numbers of qubits.
\subsection{Impact of Noise}
\begin{figure*}
    \centering
    \includegraphics[width=\linewidth, keepaspectratio]{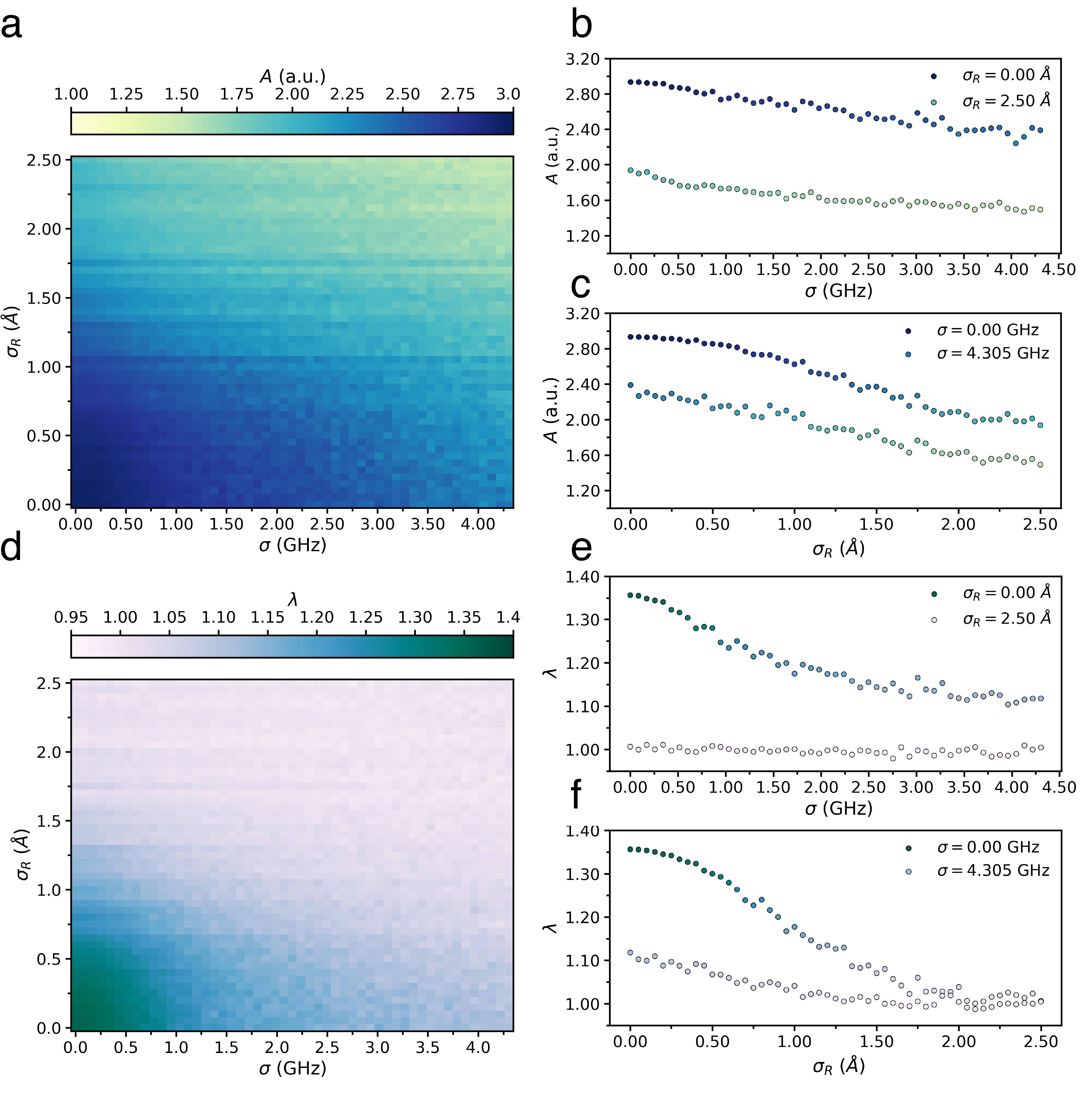}
    \caption{Heat plots of $A$ (a) and associated $\lambda$ value (c) for a system of three spin qubits, with Gaussian noise added according to Eq. \ref{eq:Hinoise} and Eq. \ref{eq:Hidipnoise} for increasing standard deviations $\sigma$ (increasing along the x axis of the plots) and $\sigma_\mathrm{R}$ (increasing along the y axis of the plots). Each data point is the average of 100 repeated calculations. The $A$ values for increasing $\sigma$ at constant $\sigma_\mathrm{R}$ values of 0.00 {\AA}  and 2.50 {\AA} are plotted in (b), with the corresponding $\lambda$ values plotted in (e). The $A$ values for increasing $\sigma_\mathrm{R}$ at constant $\sigma$ values of 0.00 GHz  and 4.305 GHz are plotted in (c), with the corresponding $\lambda$ values plotted in (f).}
    \label{fig:noisy}
\end{figure*}

The calculations discussed thus far have considered pristine arrays of spin qubits with ideal, identical local environments. However, any experimentally prepared system would include deviations from these idealized parameters. As such, it is relevant to consider the effects of such deviations, or ``noise", on the phenomena of an enhanced collective state. One source of noise expected to be relevant in an experimental context is deviations from the zero-field splitting D parameter, which can be caused by variations in temperature, pressure, and the lattice environment of the spin.\cite{ivady_pressure_2014, cambria_physically_2023, jani_optically_2024}  To consider the effects of these variations, we add energy fluctuations as Gaussian random noise to each of the individual center Hamiltonians $\op{H}^\mathrm{\mathrm{(i)}}$ in Eq.~\ref{eq:Htot}:
\begin{equation}
\op{H}^\mathrm{\mathrm{(i)}} = (D+\delta D)(\op{S}_\mathrm{Z}^2-\frac{1}{3}\op{S}^2),
\label{eq:Hinoise}
\end{equation}
where $\delta D$ are Gaussian random variables with standard deviation $\sigma$ added to the zero-field splitting D parameter of each center. The dipole-dipole interaction between triplets can then be added according to Eq.~\ref{eq:Hdip}, as in the noiseless case, with maximum transition amplitude $A$ and the corresponding eigenvalue $\lambda$ calculated as discussed above. The effect of increasing amounts of noise added to the zero-field parameter can be modeled by increasing the standard deviation $\sigma$ of the Gaussian distribution. The standard deviation is scaled from zero to 1.5 times the ideal NV center D parameter, in order to capture a realistic range of possible deviations from the parameter.\cite{ivady_pressure_2014,cambria_physically_2023,jani_optically_2024}

To capture the effects of structural disorder, Gaussian random noise is also added to the position vectors of the qubits, which impacts the dipole interaction Hamiltonian as in Eq.~\ref{eq:Hidipnoise}
\begin{equation}
\begin{split}
\op{H}^\mathrm{(ij)}_\mathrm{\mathrm{dipole}} =\frac{\hbar\gamma_\mathrm{el}^2}{|R_\mathrm{ij}+\delta_\mathrm{R_{ij}}|^3}&\left[\op{S}^\mathrm{i}\cdot\op{S}^\mathrm{j}-3\left(\op{S}^\mathrm{i}\cdot\frac{\op{R}+\op{\delta}_\mathrm{R_{ij}}} {|R_\mathrm{ij}+\delta_\mathrm{R_{ij}}|}\right)\right. \\
&\left. \left(\op{S}^\mathrm{j}\cdot\frac{\op{R}_\mathrm{ij}+\op{\delta}_\mathrm{R_{ij}}}{|R_\mathrm{ij}+\delta_\mathrm{R_{ij}}|}\right)\right],
\end{split}
\label{eq:Hidipnoise}
\end{equation}
where $\delta_\mathrm{R_{ij}}$ are Gaussian random variables with standard deviation $\sigma_\mathrm{R}$ added to the position of each qubit (random variable added to the x, y, and z coordinates). We calculate $A$ and $\lambda$ for increasing amounts of noise added to the zero-field parameter by increasing the standard deviation $\sigma_\mathrm{R}$ of the Gaussian distribution from 0 to 2.5 {\AA}.

These two noisy Hamiltonians are combined to make a total noise Hamiltonian, with $A$ and $\lambda$ for increasing amounts of each type of noise added (each standard deviation scaled over the stated range). These results are provided in two 2D heatplots (Fig. \ref{fig:noisy}a and d), where each data point is the average of 100 repeated calculations. The results of these calculations demonstrate that the enhanced collective state is susceptible both to structural disorder and to distortions of the zero-field splitting parameter, with both $A$ and $\lambda$ decreasing as each type of noise is increased. While noise that impacts the zero-field splitting alone does not completely quench the collective state for the range of standard deviation $\sigma$ used here, evidenced by the $\lambda$ value never reaching the uncorrelated value of 1.0 for $\sigma_\mathrm{R}=0.0$ (Fig. \ref{fig:noisy}e, top), it does significantly diminish the amount of correlation, suggesting that large enough fluctuations in the zero-field splitting parameter could eventually quench the enhanced collective state. However, these fluctuations would have to be very large in magnitude, falling outside of the range of reasonable D parameter distortions upon which we based our standard deviations.\cite{ivady_pressure_2014,cambria_physically_2023,jani_optically_2024} The enhanced collective state can be quenched by a significant amount of structural disorder alone, with $\lambda$ reaching 1.0 for a random distortion with standard deviation of 2.0 {\AA} added to the position of each qubit in the x, y, and z directions (Fig.~\ref{fig:noisy}f, top). This is consistent with the findings discussed in earlier sections which highlight the importance of inter-qubit alignment and spacing for the existence of an enhanced collective state. In general, while large amounts of both types of noise will disrupt the strong correlation in the system and can eventually destroy the enhanced collective state, the heatplots in Fig.~\ref{fig:noisy}a and d demonstrate a region where an enhanced collective state is still present even in the presence of both types of noise. This collective state may not be as robust as in the pristine noiseless system, but entanglement enhancement is still present, evidenced by the fact that both $A$ and $\lambda$ are above the non-interacting threshold values of 1.0. We believe that this result supports the potential, even in the presence of noisy settings, for experimental preparation of an enhanced collective state of spin qubits for quantum sensing.

\section{Conclusions and Outlook}

In this work, we introduce a pathway for entanglement-enhanced multi-qubit quantum sensing based on the preparation of an enhanced collective state, defined by the existence of a transition operator with a large transition amplitude between that state and the ground state. This increased transition amplitude can lead to stronger spin-dependent optical signals. Such a state displays a particular type of entanglement, ODLRO in the particle-hole reduced density matrix, whose signature is the presence of an eigenvalue greater than one. This eigenvalue, denoted $\lambda$, is an entanglement witness used previously in the context of exciton condensation.\cite{garrod_particle-hole_1969, safaei_quantum_2018, payne_torres_molecular_2025, liu_entanglement_2025} We show that an entangled state with ODLRO can be prepared in a system of spin qubits with strong dipole interactions, with strong particle-hole entanglement signified by the emergence of a large value of $\lambda$. Such a state can be interpreted as a quantum condensate of spin qubits. Transitions into this state exhibit increased transition amplitudes, with the largest transition amplitudes occurring at geometries that maximize the large eigenvalue.

The enhanced collective state explored in this work can be viewed as one member of a broader family of collective states unified by a common feature: the presence of strong particle-hole entanglement, signaled by a large eigenvalue of the particle-hole reduced density matrix. This family includes entangled excitonic states in photosynthetic light-harvesting complexes, where highly efficient energy transfer is associated with a collective exciton delocalized across multiple chromophores.\cite{schouten_exciton-condensate-like_2023,schouten_exciton-condensate-like_2025} It also encompasses macroscopic exciton condensation, which can be understood as an extreme limit of an entangled particle-hole collective state.\cite{payne_torres_molecular_2025, Liu.2017, Liu.2022, Pannir-Sivajothi.2022} A further example is provided by molecular J-aggregates, whose enhanced transition dipoles arise from a correlated excitonic state and likewise scale as $\mathcal{O}(\sqrt{N})$.\cite{thilagam_entanglement_2011, wurthner_jaggregates_2011, deshmukh_bridging_2022} Taken together, these examples illustrate how the particle-hole entanglement witness offers a unifying lens through which to connect systems and phenomena previously regarded as unrelated.

Our results present a strategy for harnessing entanglement as a resource for high-precision quantum sensing. The realization of entangled quantum sensors has been an extremely desirable goal, with the potential to improve sensitivity beyond the standard quantum limit.\cite{huang_entanglement-enhanced_2024,zaiser_enhancing_2016,marciniak_optimal_2022} However, experimental efforts thus far have been stymied by the vulnerability of such systems to environmental noise.\cite{huang_entanglement-enhanced_2024} The type of entangled quantum sensing proposed here, which utilizes a form of particle-hole entanglement previously predicted to be resilient in noisy biological systems,\cite{schouten_exciton-condensate-like_2023,schouten_exciton-condensate-like_2025} offers a promising avenue for the practical realization of entanglement-enhanced quantum sensing. Our results highlight the sensitivity of the enhanced collective state to structural features such as inter-qubit spacing and alignment, suggesting that a high degree of controllability of these features may be desirable for the physical realization of this phenomenon. In particular, we find that the enhanced collective state is maximized when the triplet spins are spaced closely together and aligned with the axis of microwave propagation, an insight that could inform experimental studies of the phenomenon. Interactions with qubits not aligned with the axis of microwave propagation do not appear to augment the enhanced collective state, but they also do not quench it, suggesting that the phenomenon may be feasible even without a pristine one-dimensional array of qubits. Indeed, calculations that include Gaussian random noise in both the zero-field splitting and the geometry of the spin qubits demonstrate that the entanglement enhancement can persist even in the presence of environmental noise and structural disorder. Together, our results offer a design principle for enhanced sensitivity in quantum sensing, which may be applicable to newly developed quantum sensor candidates such as molecular arrays and fluorescent proteins. As these systems have the potential for high tunability, spatial resolution, and easy integration into biological environments,\cite{kultaeva_atomic-scale_2022, yamabayashi_scaling_2018, bayliss_optically_2020, feder_fluorescent-protein_2024} combining our approach with the increased sensitivity offered by enhanced collective states could lead to exciting advances for quantum sensing in biological systems.


\begin{acknowledgments}
All authors acknowledge support from the U.S. National Science Foundation (NSF) QuBBE Quantum Leap Challenge Institute (NSF Grant no. OMA-21211044).  D. M. also gratefully acknowledges support from the NSF (Grant nos. OSI-2427090 and CHE-2155082).  I.A. also gratefully acknowledges the NSF Graduate Research Fellowship Program under Grant no. 2140001.
\end{acknowledgments}

\section*{Methods}

All calculations are carried out with Python and the QuTiP quantum simulation package,\cite{johansson_qutip_2012,johansson_qutip_2013,lambert_qutip_2024} and calculation results were plotted with matplotlib.\cite{hunter_matplotlib_2007} Transition amplitudes are calculated from the eigenstates of the total Hamiltonian $\hat{H}^\mathrm{N}_\mathrm{interacting}$ with the microwave Hamiltonian $\hat{H}_\mathrm{micro}$ as the transition operator. The particle-hole RDM is calculated for the eigenstate of the total Hamiltonian with the largest transition amplitude, with elements as given by equation (\ref{Gmat}).

For our triplet spin system, the overall particle-hole RDM is composed of $N^2$ $9 \times 9$ submatrices, where each element is the expectation value of the creation and annihilation operator terms. To create the modified particle-hole RDM, we subtract the elements corresponding to the state-to-state projection:
\begin{equation}
^2\Tilde{G}^{i,j}_{k,l}={}^2G^{i,j}_{k,l}-{}^{1}D^{i}_{j}{}^{1}D^{l}_{k},
\label{eq:modGmat}
\end{equation}
where the ${}^{1}D$ terms are elements of the singe-particle RDMs. The eigenvalues and eigenvectors can then be calculated from this modified particle-hole RDM, with $\lambda$ being the largest eigenvalue.

\bibliography{references}

\end{document}